\documentstyle[epsf,emulateapj,apjfonts]{article}
\cpright{AAS}{2005}
\def\Mtot{M_{tot}}
\def\Mstel{M_{st}}
\def\re{r_e}
\def\rl{r_{lens}}
\def\nap{N05}
\slugcomment{Accepted for publication in the Astrophysical Journal Letters}
\begin{document}

\title{Stellar and total mass in early-type lensing galaxies}
\author{Ignacio Ferreras$^{1}$,
Prasenjit Saha$^2$, and Liliya L.R. Williams$^3$}
\affil{
$^1$ Department of Physics and Astronomy, University College London,
  Gower St. London WC1E 6BT, England\\
$^2$ Astronomy Unit, Queen Mary and Westfield College, University of London,
  London E1~4NS, England\\
$^3$ Dept. of Astronomy, University of Minnesota, 116 Church Street SE,
  Minneapolis MN 55455\\
}
\authoremail{ferreras@star.ucl.ac.uk}

\begin{abstract} 
For $18$ well-observed gravitationally lensed QSOs, we
compare new non-parametric mass profiles for the lensing galaxies with
stellar-population models derived from published HST photometry.
The large volume of parameter space searched -- with respect to the
possible star formation histories -- allows us to infer
robust estimates and uncertainties for the stellar masses.
The most interesting results are: (1) the transition from little or no
dark matter in the inner regions ($\lesssim r_e$) to dark matter 
dominating on the $\sim 5r_e$ scale ($\sim 20$kpc) is clearly seen 
in massive ellipticals; 
(2) Such a trend is not seen in lower-mass galaxies, so that 
the stellar content dominates the mass budget out to $\sim 5r_e$;
(3) the radial gradient in the dark-matter fraction for these
intermediate redshift galaxies agrees with published data on nearby galaxies.
This result can help reconcile the discrepancies found in recent
estimates of dark matter in elliptical galaxies using different 
techniques (e.g. Planetary nebulae versus X-ray). The observed trend
suggests the stellar component in massive galaxies extends further out
in terms of the dark matter scale radius.
\end{abstract}


\section{INTRODUCTION}

For galaxies further than $\sim100$Mpc, the usual kinematic tracers of
galactic dynamics become increasingly difficult or impossible to use.
But at larger distances, nature sometimes provides a very different
indicator of galaxy mass---strong lensing of quasars.  Galaxies
with a quasar conveniently placed behind them, which they then lens into
multiple images, are rare. But for those
galaxies it is fairly easy to measure masses, even to $z\sim1$.  In
recent years there has been progress on estimating the $M/L$ in samples
of lensing galaxies (Keeton et al.\ 1998, Kochanek et al.\ 2000, Rusin
et al.~2003).  For a few systems there have been efforts to combine
lensing and velocity dispersions (Treu \& Koopmans 2002a,b, Koopmans \&
Treu 2003).

In this Letter we go beyond simple $M/L$ for a sample of lensing
galaxies and try to recover the distribution of stellar and dark mass
within galaxies.  Our technical innovations are (i)~we use observed
colors to model the stellar population in detail and hence map the
stellar mass, and (ii)~we use the method of pixelated lens
reconstruction to make detailed profiles of the total mass; we pay
particular attention to quantifying the uncertainties in both
departments.

Our sample comprises 17 early-type galaxies over a wide range of redshifts
($0.3<z<1$) and a bulge ($z=0.04$). The sample has been selected from 
the CASTLeS group database\footnote{See the CASTLeS page, 
{\tt http://cfa-www.harvard.edu/glensdata/} for a list.}. The main
properties are listed in Table~1.
We assume a concordance cosmology ($\Omega_{\rm
m}=0.3,\Omega_\Lambda=0.7$) with $H_0^{-1}=14\rm\,Gyr$.

\begin{figure*}[tb]
    \parbox[t]{3.2in}{
	\begin{center}
	\leavevmode
	\epsfxsize = 3.1in 
	\epsfbox{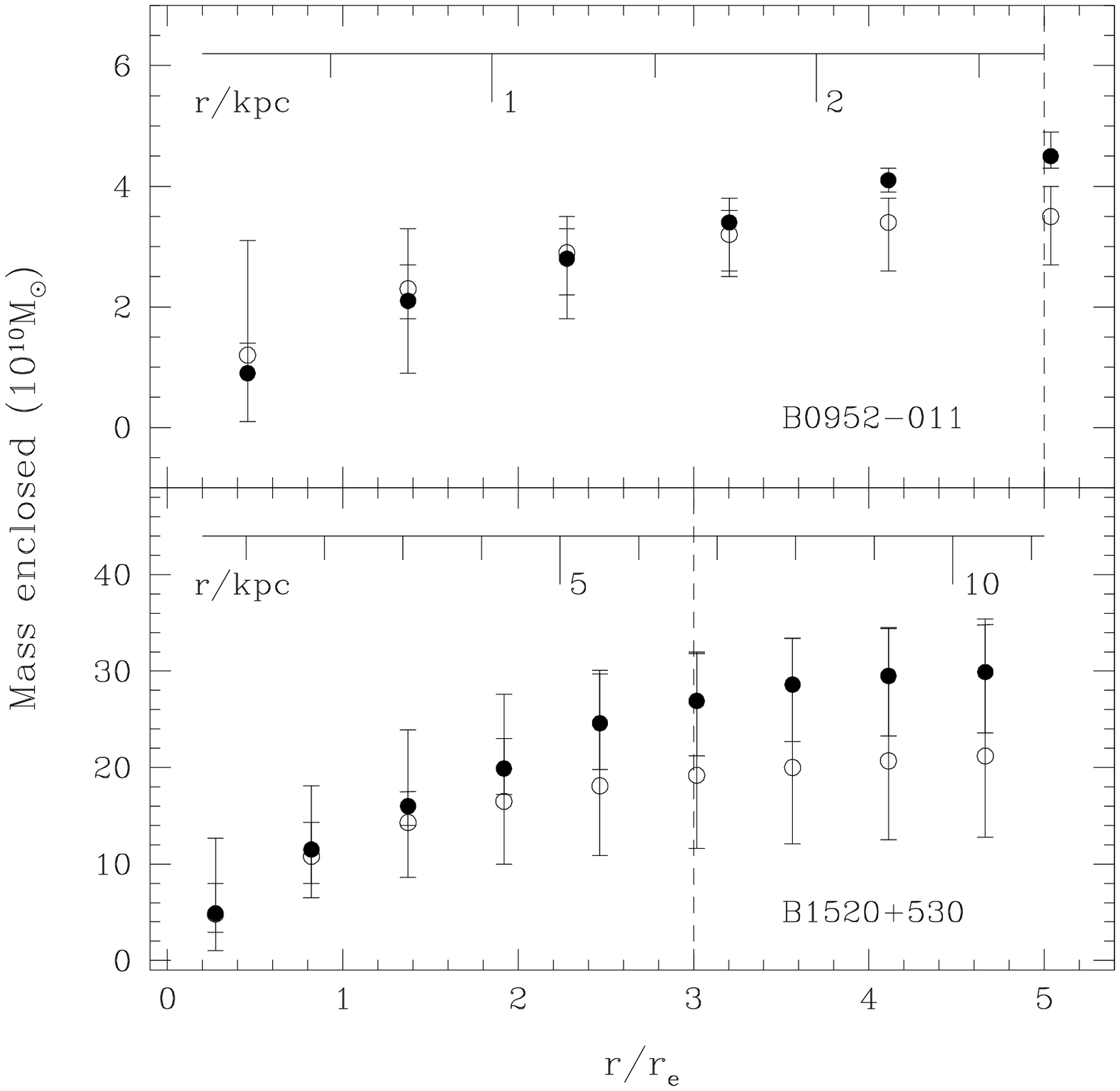}
	\end{center}
    }	
    \hfill
    \parbox[t]{3.2in}{
	\begin{center}
	\leavevmode
	\epsfxsize = 3.1in 
	\epsfbox{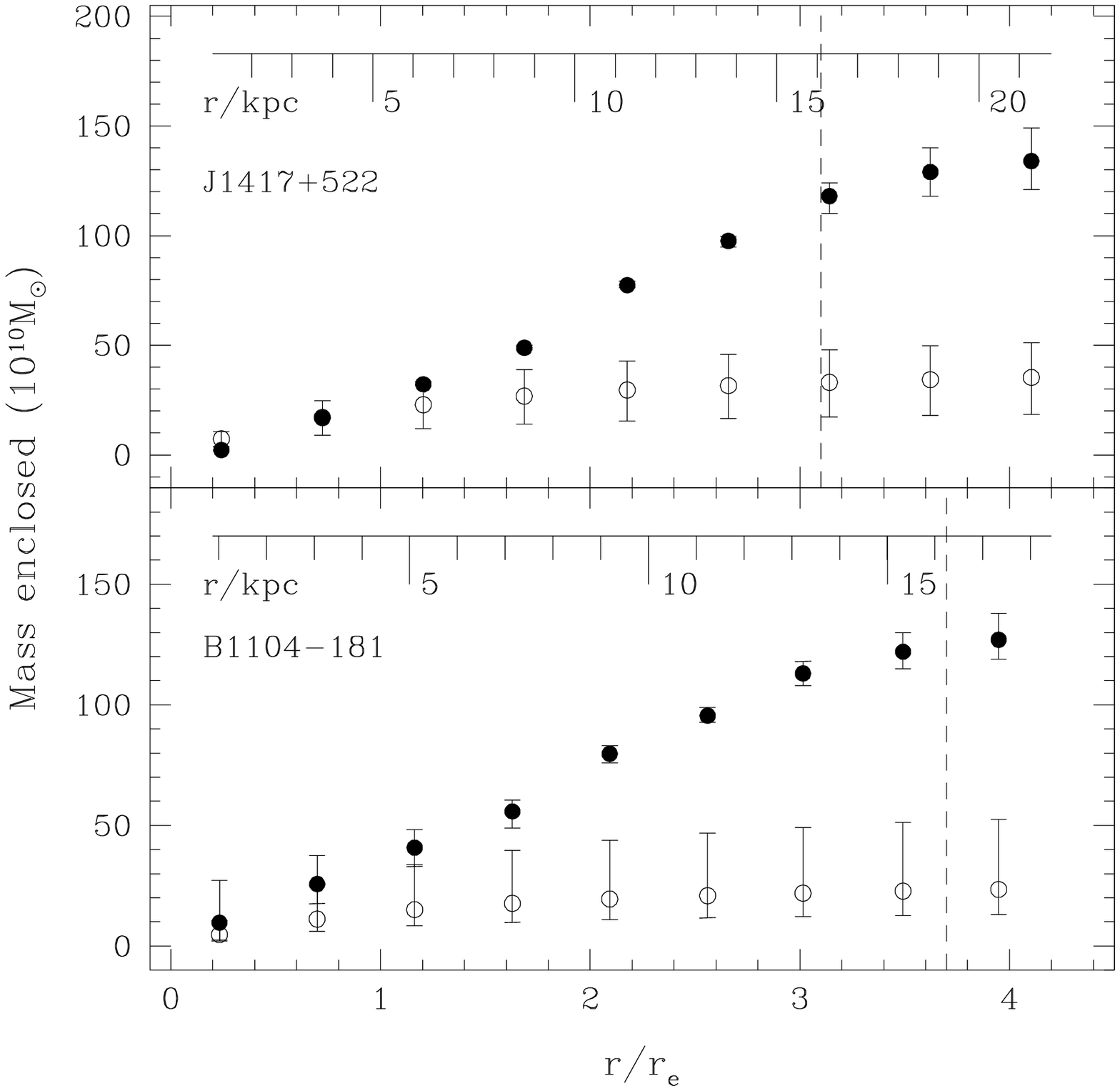}
	\end{center}
    }
    \vskip -0.3in
    \caption{Profiles of the mass enclosed as a function of radius for
       two low-mass (left) and two massive early-type galaxies
       (right).  The solid and empty circles give the total and
       stellar mass content, respectively. The stellar masses assume a
       Chabrier IMF but a Salpeter IMF leads to qualitatively the same
       result. The error bars indicate 90\% confidence intervals. The vertical
       dashed line gives the position of $\rl$.}
\label{fig:profile}
\vskip -0.2in
\end{figure*}

\section{Measuring lensing masses}\label{totalmass}

For most lensed quasars, it is fairly easy to fit a model galaxy lens to
the image positions and hence provide a map of the total sky-projected
mass.  But such a map, although a reasonable first approximation, will
be non-unique because of lensing degeneracies (Falco et al. 1985,
Gorenstein et al. 1988, Saha 2000).  A way around this problem is to
generate an ensemble of models (Williams \& Saha 2000, Trotter et al.\
2000, Keeton \& Winn 2003) all models constrained to reproduce the
observed image positions (and time delays if known) precisely. From the
model ensemble, estimates and uncertainties of any desired
quantity---for example, the mass at a given projected radius---can be
easily extracted.  We use the {\em PixeLens\/} code (Saha \& Williams
2004) which more or less automates the whole procedure, even for complex
lenses like B1608+656.

The model-ensemble technique (we used 200 models per lens) makes our
mass uncertainties larger than in previous work on $M/L$ ratios, but
much more realistic.  Four-image lenses tend to be better constrained
than two-image lenses. Also, systems with known time delays allow much
tighter mass estimates (provided $H_0$ is assumed, as we do here).

\section{Measuring stellar masses}

The stellar mass content can be determined from the photometry although
a fair share of assumptions must be invoked in order to transform light
into mass.  These assumptions relate to the age, metallicity and mass
distribution of the unresolved stellar populations.  The latter can be
reduced to a time-independent, universal initial mass function (IMF) as
suggested by observations (see e.g. Wyse 1998). In this paper we explore two
IMFs: Salpeter (1955) and Chabrier (2003), both defined between
$0.1M_\odot$ and $100M_\odot$. The former assumes a simple
power law over the allowed range of stellar masses. The IMF proposed by
Chabrier (2003) has a very similar upper-mass dependence as the Salpeter
function. However, the low-mass end assumes a flatter -- more physical
-- behaviour, following a lognormal distribution. This IMF gives $M/L$
ratios which are a factor $\sim 1.5$ smaller than those for a Salpeter
IMF.

The main difference between most of the IMFs proposed
(e.g. Salpeter 1955; Miller \& Scalo 1979; 
Kroupa, Tout \& Gilmore 1993; Chabrier 2003) lies in the relative
contribution from stars with masses $M< M_\odot$. 
Since these stars have a large $M/L$ ratio, stellar populations 
born with either IMF feature very similar spectral energy 
distributions, albeit with a different normalization for a given 
stellar mass (see e.g., Bruzual \& Charlot 2003). 
Therefore, a photo-spectroscopic analysis cannot be
used reliably to constrain the IMF at the low-mass end.
Instead, we will use the available photometry from our sample and
compute the stellar mass content corresponding to either a Salpeter
or a Chabrier IMF. These two mass distributions
represent a robust range of possible values between the
unphysically high $M/L$ imposed by a simple power law extrapolated
to low masses and a more realistic distribution.

The age and metallicity distribution can be constrained by a comparison
of the photometry with a simple model of star formation. We assume a
3-parameter model which reduces the description of the stellar
populations to a single metallicity ($Z_\star$), 
a formation epoch ($t_{\rm FOR}$), and a formation timescale ($\tau_{\rm
SF}$), so that at any given time the star formation rate is
$\psi(t)\propto\exp (-\Delta t/\tau_{\rm SF})$, with $\Delta t=t-t_{\rm
FOR}$. Each choice of parameters $(Z_\star ,t_{\rm FOR},\tau_{\rm SF})$
represents a possible formation scenario, and can be convolved with
simple stellar populations (SSP) in order to generate a composite model
from which various photo-spectroscopic observables can be retrieved. We
use the Bruzual \& Charlot (2003) population synthesis models.

We use the published photometry of the lenses from the CASTLeS group
(Rusin et al. 2003) which was obtained from HST/WFPC2+NICMOS images
after carefully subtracting the contribution from the sources. Each set
of measurements correspond to a number of colours (from $1$ to $5$)
which is used in order to constrain the parameter space described above.
We use their radial fits to generate the profile of 
the stellar component.
Galactic reddening was included in the analysis by using the $E(B-V)$
values from Rusin et al. (2003) and applying a dust correction according
to the R=3.1 curve of Fitzpatrick (1999).

The best fit for each lens was obtained using an adaptive grid in the
3-dimensional parameter space that describes all possible star formation
histories in our model, and using a Metropolis algorithm (see e.g.,
Binney et al. 1992, or Saha 2003) to find the uncertainties which are
quoted with respect to the 5th-95th percentiles (90\% confidence
interval) throughout the paper.  The stellar mass content is computed
for the best fit, with respect to the light inside the half-light radius
($r_e$).

\begin{figure*}[tb]
    \parbox[t]{3.5in}{
	\begin{center}
	\leavevmode
	\epsfxsize = 3.4in 
	\epsfbox{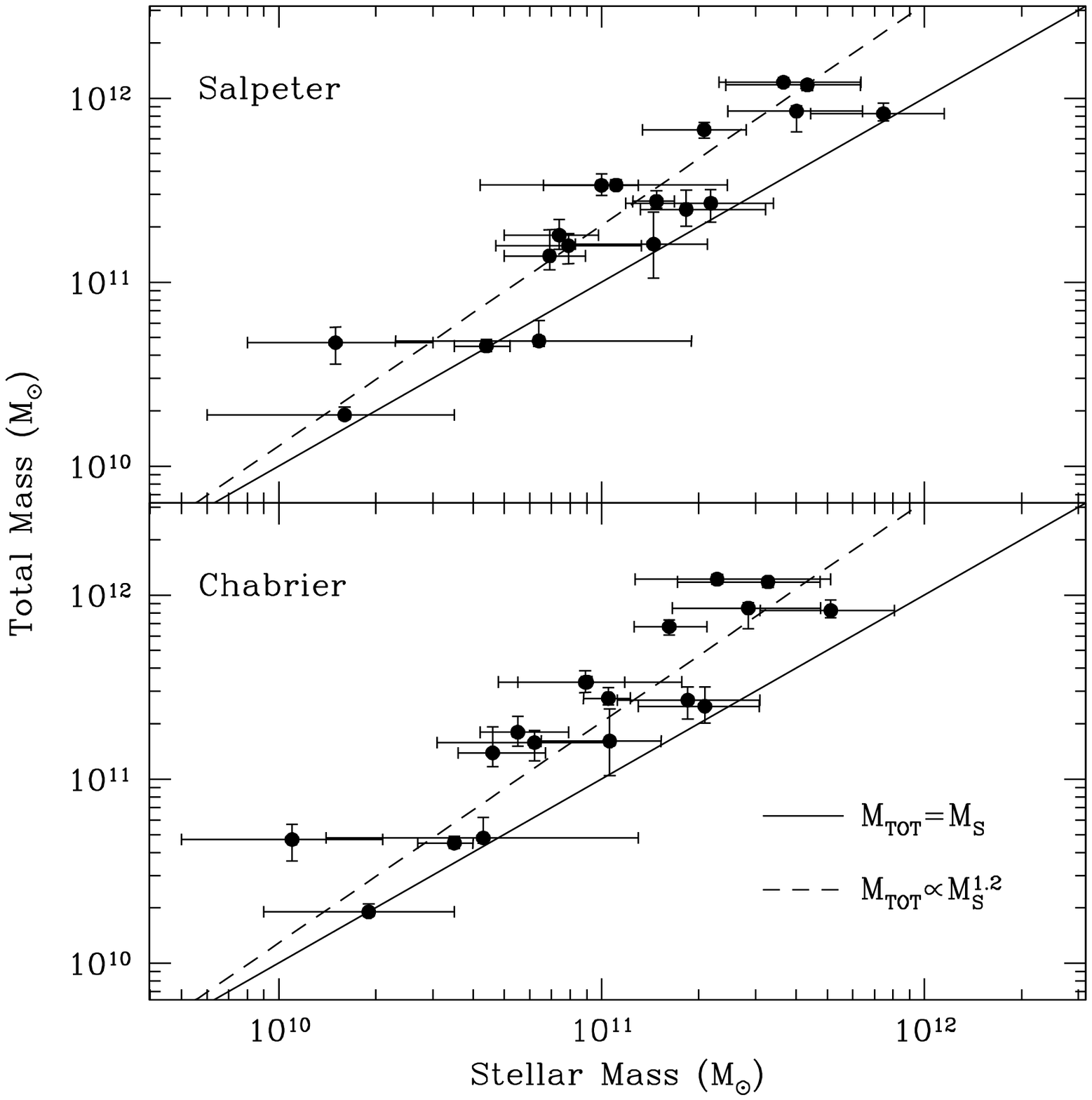}
	\vskip -0.4in 
	\end{center}
	\caption{A comparison of total and stellar masses for our
	  sample of gravitational lens early-type galaxies 
	  (all measured inside $\rl$). The upper
	  and lower panels correspond to a Salpeter and a Chabrier
	  IMF, respectively.  The solid line represents $\Mtot/\Mstel$
	  and the dashed line follows the expected correlation from
	  the tilt of the fundamental plane for ellipticals.}
	\label{fig:mass}
    }	
    \hfill
    \parbox[t]{3.5in}{
	\begin{center}
	\leavevmode
	\epsfxsize = 3.1in 
	\epsfbox{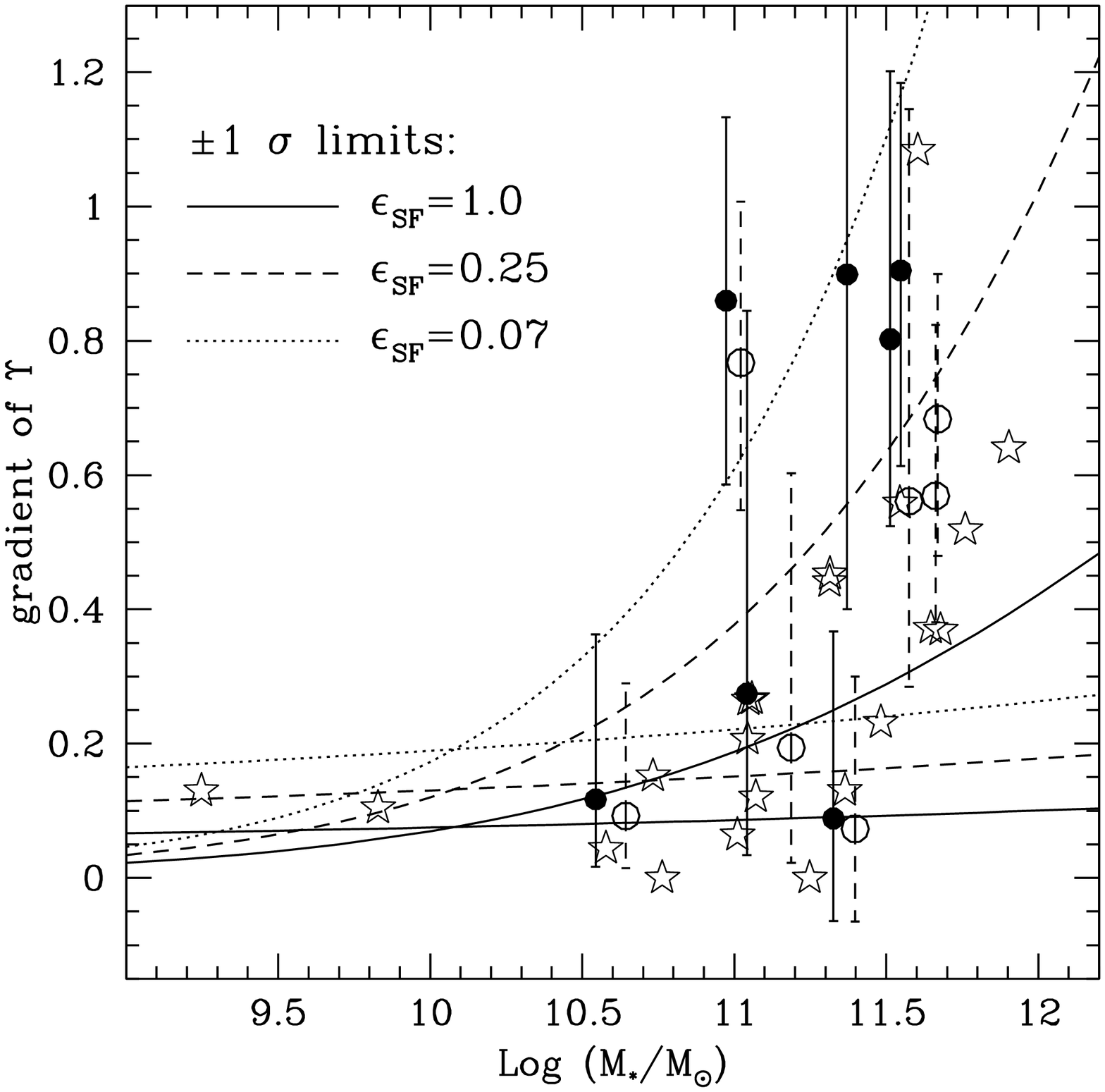}
	\end{center}
        \caption{Dark matter gradient $\nabla\Upsilon$, computed between
          $r_{in}\simeq 0.25\re$ and $r_{out}\simeq 4.5\re$.  Filled circles
          correspond to the Chabrier IMF and empty circles to the
          Salpeter IMF.  The error bars correspond roughly to $1\sigma$.
          This figure may be compared with Fig.~7 of Napolitano et al.\
          (2005).  The star symbols are taken from that paper.  The
          curves are $1\sigma$ bands for model predictions with
          different star-formation efficiencies, also adapted from
          Napolitano et al.}
        \label{fig:MLgrad}
}
\vskip -0.2in

\end{figure*}

\section{Stellar vs Total Mass}

Figure \ref{fig:profile} shows the mass profiles (stellar and total) for
four galaxies, two with $\Mtot \ll 10^{12}M_\odot$ ({\sl left}) and two with
$\Mtot\gtrsim 10^{12}M_\odot$ ({\sl right}). 
The result is striking: the low-mass galaxies
have little or no dark matter at all observed radii.  
The high-mass galaxies have little or no
dark matter inside of $r\sim\re$, but at large radii they are dominated
by dark matter.  Here we have a clear indication that massive galaxies
have a gradient in their dark-matter fraction.

Rather than presenting mass profiles for all the galaxies in our
sample, we will summarize the profiles in two ways: first by comparing
stellar and total mass at a chosen radius, and then by quantifying the
dark-matter gradient.

In Figure~\ref{fig:mass} and Table~\ref{tab1} we show three derived
aperture masses: the total mass derived from lensing, the stellar mass
derived assuming a Salpeter IMF, and the stellar mass derived assuming
a Chabrier IMF. Uncertainties are at 90\% confidence.  The aperture
radius in all cases is a quantity we call $\rl$, and it is the
largest radius at which the lensing mass is well-constrained.  In
general, $\rl$ represents the radius over which there is lensing
information, and is roughly the radius of the outermost image.  From
Figure~\ref{fig:mass} it is apparent that the dark-matter fraction
increases with mass, and the trend is compatible with the observed
tilt of the NIR fundamental plane (Mobasher et al.~1999; Ferreras \&
Silk 2000). Additionally, we see that $\Mstel$ is estimated as more
than $\Mtot$ in a few cases for the Salpeter IMF, but never for the
Chabrier IMF; while it would be cavalier to argue against a Salpeter
IMF on this basis, it does suggest that future comparisons of lensing
and stellar-population models may help constrain the IMF.

We now consider the gradient in the dark-matter fraction. Following
Napolitano et al.\ (2005; hereafter \nap) we introduce
\begin{equation}
\nabla\Upsilon \equiv {r_e\over\Delta r}
      \left[ \left(\Mtot\over\Mstel\right)_{\rm out} -
      \left(\Mtot\over\Mstel\right)_{\rm in} \right] .
\end{equation}
Even though $\nabla\Upsilon$ is a noisy quantity, it has advantages over
$\Mtot/\Mstel$.  Being a gradient it is not sensitive to the choice of
wavebands used to measure stellar luminosity. Also, $\nabla\Upsilon$ is
unlikely to be very sensitive to the redshift evolution of galaxies,
because to first order $\Mtot/\Mstel$ will change at the same rate
throughout the inner parts of a galaxy as its stellar population ages.
We are able to compute $\nabla\Upsilon$ with tolerable uncertainty for
only a subsample of galaxies. 

The work of \nap\  differs from ours in several respects: (i)~they
derived masses from the kinematics of planetary nebulae and globular
clusters; (ii)~their galaxies are nearby whereas ours span range from
$z\simeq0.3$ to and $z\simeq1$; (iii)~they used three-dimensional radial
gradients while we are working with projected quantities.  Point (ii) is
unlikely to affect $\nabla\Upsilon$ much, as we have already argued.
Regarding point (iii), the difference is minimal for the estimation of
the stellar masses, since as Figure~\ref{fig:profile} shows, the stellar
mass is concentrated within the central few kpc; even for the more
extended distribution of the dark matter, most of the mass that makes up
the total line of sight mass resides at small radii. In general, 2D and 3D
$\nabla\Upsilon$ estimates differ by less than 50\%. We conclude that
we can directly compare with \nap.

{\small 
\begin{deluxetable}{lrrcrccc}
\tablewidth{13cm}
\tablecaption{Aperture Mass ($\times 10^{10}M_\odot$)\label{tab1}}
\tablehead{\colhead{Object} & \colhead{z} & \colhead{r$_e$/kpc} & 
\colhead{r$_{lens}$/kpc} & \colhead{$M_V$} & \colhead{M$_{tot}(<\rl )$} 
& \colhead{M$_{salp}(<\rl )$} & \colhead{M$_{chab}(<\rl )$}}
\startdata
B0047-280 & 0.49 &  5.4 &  8.6 & $-$22.5 & ${33.7}_{32.3}^{36.2}$
          & ${11.1}_{ 4.2}^{24.5}$ & ${ 9.0}_{ 4.8}^{17.7}$ \nl
B0142-100 & 0.49 &  3.1 &  5.6 & $-$22.7 & ${24.9}_{20.2}^{31.7}$
          & ${18.3}_{13.2}^{32.2}$ & ${20.9}_{13.0}^{30.8}$ \nl
J0414+053 & 0.96 & 14.5 & 29.0 & $-$23.1 & ${82.5}_{75.3}^{93.9}$
          & ${74.7}_{44.5}^{115.4}$ & ${51.3}_{31.0}^{80.8}$ \nl
B0818+122 & 0.39 &  4.8 & 10.6 & $-$22.0 & ${67.4}_{60.7}^{73.6}$
          & ${20.8}_{13.4}^{28.1}$ & ${16.2}_{12.6}^{21.2}$ \nl
J0951+263 & 0.20 &  0.7 &  2.1 & $-$19.3 & ${ 4.7}_{ 3.6}^{ 5.7}$
          & ${ 1.5}_{ 0.8}^{ 3.0}$ & ${ 1.1}_{ 0.5}^{ 2.1}$ \nl
B0952-011 & 0.38 &  0.5 &  2.5 & $-$19.4 & ${ 4.5}_{ 4.2}^{ 4.9}$
          & ${ 4.4}_{ 3.5}^{ 5.2}$ & ${ 3.5}_{ 2.7}^{ 4.0}$ \nl
B1009-025 & 0.78 &  1.6 &  4.8 & $-$21.0 & ${18.0}_{15.1}^{22.0}$
          & ${ 7.4}_{ 5.0}^{ 9.8}$ & ${ 5.5}_{ 4.2}^{ 7.9}$ \nl
J1017-204 & 0.86 &  2.4 &  2.4 & $-$21.6 & ${ 4.8}_{ 4.5}^{ 6.2}$
          & ${ 6.4}_{ 2.3}^{19.0}$ & ${ 4.3}_{ 1.4}^{13.0}$ \nl
B1030+074 & 0.60 &  2.6 &  4.7 & $-$21.8 & ${16.1}_{10.5}^{24.1}$
          & ${14.5}_{ 8.3}^{21.3}$ & ${10.6}_{ 6.5}^{15.3}$ \nl
B1104-181 & 0.73 &  4.4 & 16.3 & $-$22.7 & ${122.0}_{115.0}^{130.0}$
          & ${36.6}_{23.1}^{63.7}$ & ${22.8}_{12.7}^{51.2}$ \nl
B1115+080 & 0.31 &  2.3 &  7.4 & $-$21.1 & ${27.5}_{25.4}^{31.5}$
          & ${14.8}_{12.5}^{16.8}$ & ${10.5}_{ 8.8}^{12.3}$ \nl
J1411+521 & 0.46 &  2.9 &  9.3 & $-$21.2 & ${33.6}_{29.6}^{38.8}$
          & ${10.0}_{ 6.6}^{13.0}$ & ${ 8.9}_{ 5.5}^{11.8}$ \nl
J1417+522 & 0.81 &  5.2 & 16.1 & $-$23.4 & ${118.0}_{110.0}^{124.0}$
          & ${43.4}_{24.3}^{63.4}$ & ${32.8}_{17.2}^{47.6}$ \nl
B1422+231 & 0.34 &  1.5 &  6.0 & $-$20.7 & ${15.8}_{12.6}^{18.4}$
          & ${ 7.9}_{ 4.7}^{13.3}$ & ${ 6.2}_{ 3.1}^{10.6}$ \nl
B1520+530 & 0.72 &  2.2 &  6.6 & $-$22.4 & ${26.9}_{21.2}^{31.8}$
          & ${21.8}_{11.9}^{34.1}$ & ${18.5}_{11.2}^{30.9}$ \nl
B1608+656 & 0.63 &  4.2 &  6.3 & $-$23.1 & ${85.0}_{65.5}^{90.9}$
          & ${40.2}_{24.6}^{64.3}$ & ${28.5}_{16.6}^{47.8}$ \nl
B2149-274 & 0.50 &  2.9 &  4.1 & $-$21.9 & ${13.9}_{11.7}^{19.3}$
          & ${ 6.9}_{ 5.0}^{ 8.9}$ & ${ 4.6}_{ 3.6}^{ 6.7}$ \nl
B2237+030 & 0.04 &  3.1 &  0.9 & $-$20.9 & ${ 1.9}_{ 1.8}^{ 2.1}$
          & ${ 1.6}_{ 0.6}^{ 3.5}$ & ${ 1.9}_{ 0.9}^{ 3.5}$ \nl
\enddata
\end{deluxetable}
}

Figure~\ref{fig:MLgrad} compares our estimates of $\nabla\Upsilon$ with
those of \nap\ in the format of their Fig.~7.  There is a general
agreement between the two sets of results, at least over the mass range
covered by both data sets. Our values are somewhat higher for
$11<\log(M_\star/M_\odot)<11.5$ range, but with the current small
numbers we cannot assess significance.  In Figure~\ref{fig:MLgrad} we
have also plotted the predictions of $\Lambda$CDM models by \nap.  This
requires some justification.  First, the models were computed for the
$\re$-$\Mstel$ relation characteristic of the \nap\ data set, but our
$\re$-$\Mstel$ turns out to be very similar.  Second, the models are
based on NFW $\Lambda$CDM predictions for halos at $z\approx 0$; do they
still apply to our data, with $z\sim0.3-1$?  Fortunately, yes, because
the relation between the concentration parameter and halo mass 
is fairly robust with respect to
the redshift of observation in the $z=0-3$ range (see Fig.~14 of
Wechsler et al.\ 2002).  The curves in Figure~\ref{fig:MLgrad} show
three different star-formation efficiencies: $\epsilon_{SF}=0.07, 0.25,$
and 1, where $\epsilon_{SF}$ is defined as the ratio between total
stellar and baryonic matter.
The general trend is that higher stellar mass systems should
have a monotonically larger contribution from dark matter with
increasing radius.
This can be understood
intuitively as follows (cf.\ section 3.4 of \nap). In galaxies with
higher stellar masses the stellar component extends to larger radii, in
terms of NFW profile scale radius, and so larger radii are more dark
matter dominated. Our data is consistent with that trend. 
However, one should be cautious when interpreting the results with
respect to $\epsilon_{SF}$. The \nap\  models disregard the effect
of the gas collapsing to the centre of the halo and its effect on the 
dark matter density profile. 

\section{Discussion}

In the Milky Way it is generally held that in the inner few kpc and on
mass scales of $\sim10^{11}M_\odot$ there is little or no dark matter,
whereas on the scale of $\sim100$kpc and $\sim10^{12}M_\odot$ dark
matter completely dominates.  In this Letter, we see the same transition
in elliptical galaxies at redshifts of $0.3$ to 1, based on projected
mass distribution derived from lensing. The dark matter 
fraction tends to increase with the total mass. 
Incidentally we note that one of the possible mechanisms that can
be invoked to explain the tilt of the fundamental plane involves such a 
trend between stellar and total mass with a very similar scaling
behavior (Ferreras \& Silk 2000). Part of the tilt is 
explained by non-homology effects (Trujillo et al. 2004), but our data
suggest a further contribution to the tilt from the different 
stellar vs dark matter distributions. We claim that
most of the tilt of the fundamental plane -- at least when the observables 
used to define it reach beyond $r_e$ -- is caused by the correlation
between dark matter fraction and galaxy mass as presented in this Letter.
The gradient of the dark matter
fraction inside individual galaxies is also similar to what is inferred
in nearby galaxies.  These results indicate that the distribution of
dark matter in galaxies has not changed much since $z=1$.

Enlarging the sample of galaxies for which stellar and dark-matter
profiles can be reconstructed in this way is an obvious goal for future
work.  But another line of future work is to improve the spatial
resolution in some cases.  A possible outcome, hinted at in this work,
is that bottom-heavy IMFs such as the Salpeter function might be 
ruled out by the comparison between stellar and lensing masses.

The correlation presented in this Letter help to reconcile the
apparent discrepancy between the low dark matter content from
dynamical estimates using planetary nebulae (Romanowsky et al.\
2003)\footnote{Recently Dekel et al.\ (2005) have argued that the
assumption of isotropic PNe orbits made by Romanowsky et al.\ (2003)
resulted in the mass being underestimated.  However, our result that
$\nabla\Upsilon$ is small for $\log (M_{st}/M_\odot)=$11--11.5
suggests that while in general velocity anisotropy is an important
effect, it is less so for a dynamically old population like PNe.}  and the
dark matter dominated galaxies suggested by X-ray data (Loewenstein \&
Mushotzky 2002).  Mass estimates of faint X-ray ellipticals, albeit
challenging, will definitively test the observed correlation.

\acknowledgments
We would like to thank Nicola Napolitano for a very useful discussion
and for providing his data and LCDM predictions.


\begin{references}
\reference{fp04} Trujillo, I., Burkert, A. \& Bell., E.~F., 
2004, ApJ, 600, L39

\reference{cp92}
Binney, J.~J., Dowrick, N.~J., Fisher, A.~J., \& Newman, M.~E.~J., 1992,
{\em The Theory of Critical Phenomena,} Oxford

\reference{bc03}
Bruzual, G. \& Charlot, S. 2003, MNRAS, 344, 1000

\reference{chabrier03}
Chabrier, G., 2003, PASP, 115, 763

\reference{dek05}
Dekel, A., et al., 2005, astro-ph/0501622 

\reference{falco85}
Falco, E.~E., Gorenstein, M.~V., Shapiro, I.~I., ApJL, 289, 1

\reference{fs00}
Ferreras, I. \& Silk, J., 2000, MNRAS, 316, 786

\reference{fitzpatrick99}
Fitzpatrick, E.~L., 1999, PASP, 111, 63

\reference{gorenstein88}
Gorenstein, M.~V., Falco, E.~E., Shapiro, I.~I., ApJ, 327, 693

\reference{kroupa93}
Kroupa, P., Tout, C. A. \& Gilmore, G., 1993, MNRAS, 262, 545

\reference{keeton98}
Keeton, C.~R., Kochanek, C.~S., Falco, E.~E. 1998, 509, 561

\reference{keeton03}
Keeton, C. R. \& Winn, J. N. 2003, ApJ, 590, 39

\reference{kochanek00}
Kochanek, C. S., et al. 2000, ApJ, 543, 131

\reference{koopmans03}
Koopmans, L. V. E., \& Treu, T. 2003, ApJ, 583, 60

\reference{lm03}
Loewenstein, M. \& Mushotzky, R., 2002, astro-ph/0208090

\reference{miller79}
Miller, G. E. \& Scalo, J. M., 1979, ApJS, 41, 513

\reference{mob99}
Mobasher, B., et al. 1999, MNRAS, 304, 225

\reference{nap05}
Napolitano, N.~R. et al., 2005, MNRAS, 357, 691

\reference{rom03}
Romanowsky, A.~J., et al., 2003, Science, 301, 1696

\reference{rusin03}
Rusin, D. et al. 2003, ApJ, 587, 143

\reference{s00}
Saha, P. 2000, AJ, 120, 1654

\reference{pda}
Saha, P. 2003, {\em Principles of Data Analysis,} Cappella Archive.

\reference{sw04}
Saha, P. \& Williams, L.~L.~R., 2004, AJ, 127, 2604

\reference{salpeter55}
Salpeter, E.~E., 1955, ApJ, 121, 161

\reference{treu02a}
Treu, T., \& Koopmans, L. V. E. 2002a, ApJ, 575, 87

\reference{treu02a}
Treu, T., \& Koopmans, L. V. E. 2002b, MNRAS, 337, L6

\reference{trotter00}
Trotter, C.S., Winn, J.N., Hewitt, J.N. 2000, ApJ, 535, 671

\reference{wmap03}
Spergel, D.~N., et al., 2003, ApJS, 148, 175

\reference{wech02}
Wechsler, R.~H., Bullock, J.~S., Primack, J.~R., Kravtsov, A.~V. \&
Dekel, A., 2002, ApJ, 568, 52

\reference{ws00}
Williams, L.~L.~R. \& Saha, P., 2000, ApJ, 119, 439

\reference{wyse98}
Wyse, R.~F.~G., 1998, in: Gilmore, G., Howell, D. (Eds.),
  The Stellar Initial Mass Function, ASP Conf. Ser., Vol. 142, 
  ASP, San Francisco, CA, p. 89
\end{references}
\end{document}